\documentclass{cjaa}                   

\usepackage{graphicx}                   
\input{epsf.sty}                        
\input{psfig.sty}                       

\begin{document}

   \title{QSO-Galaxy Association and Gravitational Lensing}



   \volnopage{Vol.0 (200x) No.0, 000--000}      
   \setcounter{page}{1}          

   \author{S. M. Tang
      \inst{1}\mailto{}
   \and S. N. Zhang
      \inst{1,2}
     }
   \offprints{}                   

   \institute{Physics Department and Center for Astrophysics, Tsinghua Univ., Beijing 100084, China\\
             \email{tangsm99@mails.tsinghua.edu.cn}
        \and
             Institute of High Energy Physics, Chinese Academic of Science, Beijing, 100039 China\\
             }

   \date{Received~~; accepted~~}

   \abstract{The amplification caused by gravitational light bending by compact objects in a foreground galaxy can affect
   the apparent number density of background QSOs, as well as their
distribution in the fields of galactic halos. In this work we investigate the number enhancement of QSOs in the
fields of galactic halos caused by point mass lensing effect and singular isothermal lensing effect, and apply the
microlensing effect due to dark compact objects in the halo to NGC 3628. NGC 3628 is a well-studied nearby edge-on
Sbc peculiar galaxy, where QSOs are shown to be concentrated around the galaxy with a density much higher than
background. We show that if present understanding of the luminosity function of QSOs is right, such concentration
could not be caused by gravitational lensing.
}

   \authorrunning{S. M. Tang \& S. N. Zhang}            
   \titlerunning{QSO-Galaxy Association and Gravitational Lensing}  

   \maketitle

%
%
\section{Introduction}           
\label{sect:intro}
Ultra-Luminous X-ray sources (ULXs) have been found from several nearby spiral galaxies and they are commonly
believed to be either stellar mass compact systems or intermediate-mass black holes (IMBHs) (see for example,
Miller and Colbert 2004 and references therein). However, a considerable fraction of these ULXs have been
identified as QSOs with high redshifts (Arp et al. 2002, 2004), and these QSOs are shown to be highly concentrated
around these galaxies; it has been proposed that these ULXs are ``local" QSOs physically associated with these
galaxies, with high intrinsic redshifts in the process of being ejected from those galaxies (Burbidge et al.
2003a, 2003b). On the other hand, if these ULX-QSO associations are real and the QSOs are cosmologically distant
objects, then there must exist a mechanism responsible for enhancing the apparent QSO number density and
concentration towards the center of their foreground galaxies in flux limited observations.

The only possible mechanism to link nearby galaxies with high redshift QSOs and to boost the number density of
QSOs in the fields of galaxies is gravitational lensing. In this paper we investigate on the question: What will
happen to the number density enhancement, if these QSOs are magnified by gravitational lensing effect due to
compact objects in galactic halos of their foreground galaxies? We find that the local fractional number density
enhancement is insignificant unless that a considerable portion of dark matter in galactic halos are compact
objects, by comparing microlenses and singular isothermal lenses with the same total mass as shown in section 2.

In section 3, we apply this microlensing effect to NGC 3628 which is a well-studied nearby edge-on Sbc peculiar
galaxy and find that the concentration of QSOs in the fields of NGC 3628 could not be caused by gravitational
lensing. Throughout this paper, we adopt a QSO luminosity function given by Ueda et al. (2003) with a 2-10 keV
flux limit of $3\times10^{-14}$ erg cm$^{-2}$ s$^{-1}$ and isotropic QSO luminosity lower limit of 10$^{44}$ erg
s$^{-1}$.

\section{QSO distribution in the fields of galactic halos}
\label{sect:galaxy-qso}

For a given total mass, galaxies at different distances will show
different gravitational lensing effects and therefore the QSO
distribution in the fields of galactic halos will be different.

\subsection{Microlenses vs smooth lenses}
For nearby galaxies, a halo with microlenses will amplify the
background QSOs much more effectively than a halo with the same
amount of mass but distributed smoothly, e.g., in the form of
WIMPs. We first compare an extreme example of lensing in which all
the mass in the galaxy is either in the form of a point mass
(i.e., microlens) or a smoothly distributed mass (i.e., singular
isothermal lens). For a galaxy with mass $M$ and typical size $L$,
the Einstein radius $r_{Ep}$ for the microlens is given by:
\begin{equation}
r_{Ep}=(\frac{4GMD}{c^2})^{1/2}\\,
\end{equation}
where $D=\frac{D_lD_{ls}}{D_s}$, and $D_l, D_s, D_{ls}$ are
angular distances between the lens (foreground galaxy) and
observer, the source (background QSO) and observer, QSO and
galaxy, respectively.

For a singular isothermal lens, its properties are characterized
by a one-component velocity distribution $\sigma_\parallel$. The
density and mass within a radius $r$ are,
\begin{equation}
\rho(r)=\frac{\sigma_\parallel^2}{2\pi G} \frac {1}{r^2}\\, \
M(r)=\frac {2\sigma_\parallel^2 r}{G}\\,
\end{equation}
and the Einstein radius $r_{Ei}$ is given by,
\begin{eqnarray}
r_{Ei}=\frac{4\pi\sigma_\parallel^2D}{c^2}=\frac{2 \pi GDM}{L
c^2}=\frac{\pi}{2} \frac{r_{Ep}^2}{L}\\.
\end{eqnarray}
We re-write the above equation as,
\begin{equation}
\frac{\pi}{2}r_{Ep}^2=r_{Ei} L\\.
\end{equation}

It is clear that $r_{Ei} < r_{Ep}$ for $D\leq2*10^3$ Mpc ($z \leq 0.4$) for a $10^{12} M_\odot$ galaxy with a
radius of 20 kpc. Since in the low-optical-depth limit and in the condition of point lens approximation, the
amplification of background QSO caused by microlenses depends on the total solid angle of Einstein rings and
consequently depends on the total mass of microlenses, the microlensing induced QSO number enhancement is
independent of the mass function of microlenses and widely-distributed microlenses would behave the same way as a
single object with a given total mass. Therefore we can conclude that microlenses are more effective than smoothly
distributed singular isothermal lenses for nearby galaxies. It is thus possible to use background QSO distribution
in the fields of galactic halos of nearby galaxies to distinguish between dark matter halos made of smoothly
distributed matter such as WIMPs from those made of point mass lenses.

\subsection{ QSO number enhancement in the fields of galaxies}

In the low optical depth regime, the microlensing induced QSO
number enhancement in the fields of galaxies is independent of the
mass function and spatial distribution of microlenses, and only
relies on the total mass of dark matter microlenses $M$ and
redshifts $z$ of galaxies (e.g., Paczynski 1986b). Fig. 1 shows
the net number enhancement $\delta N$ of QSOs in the field of a
galaxy within a radius 20 kpc as a function of redshift of the
galaxy due to microlensing of point lens dark matter. For an
observed QSO number $N$, the statistical uncertainty $\sigma$ is
$\sigma=\sqrt{N}$. This should be compared with the net number
enhancement caused by microlensing: $\frac{\delta
N}{\sigma}=\frac{\delta N}{\sqrt{N}}$, which is shown in Fig.2. We
see that:

 (1) The three curves, corresponding to different total
masses of microlenses in the galaxy, are very different. Therefore if a considerable portion of halo dark matter
is in the form of microlenses, statistics of background QSO distribution in the fields of galactic halos can be
used to determine the total mass of microlenses.

(2) Though the net number enhancement in nearby galaxies is far more than distant galaxies because they cover
larger sky areas, the statistical value $\frac{\delta N}{\sqrt{N}}$ is relatively flat as a function of the
redshifts of foreground galaxies.  This tells us that number counts of background QSOs in spatially resolved
fields of foreground galaxies are sensitive to the galaxy distances and thus may be used to study the cosmological
parameters such as dark energy, but the simple QSO-galaxy correlation is insensitive to galaxy distances. We will
return to the issue of QSO-galaxy correlation in the last part of the paper.

\begin{figure}[h]
  \begin{minipage}[t]{0.5\linewidth}
  \centering
  \includegraphics[width=65mm,height=60mm]{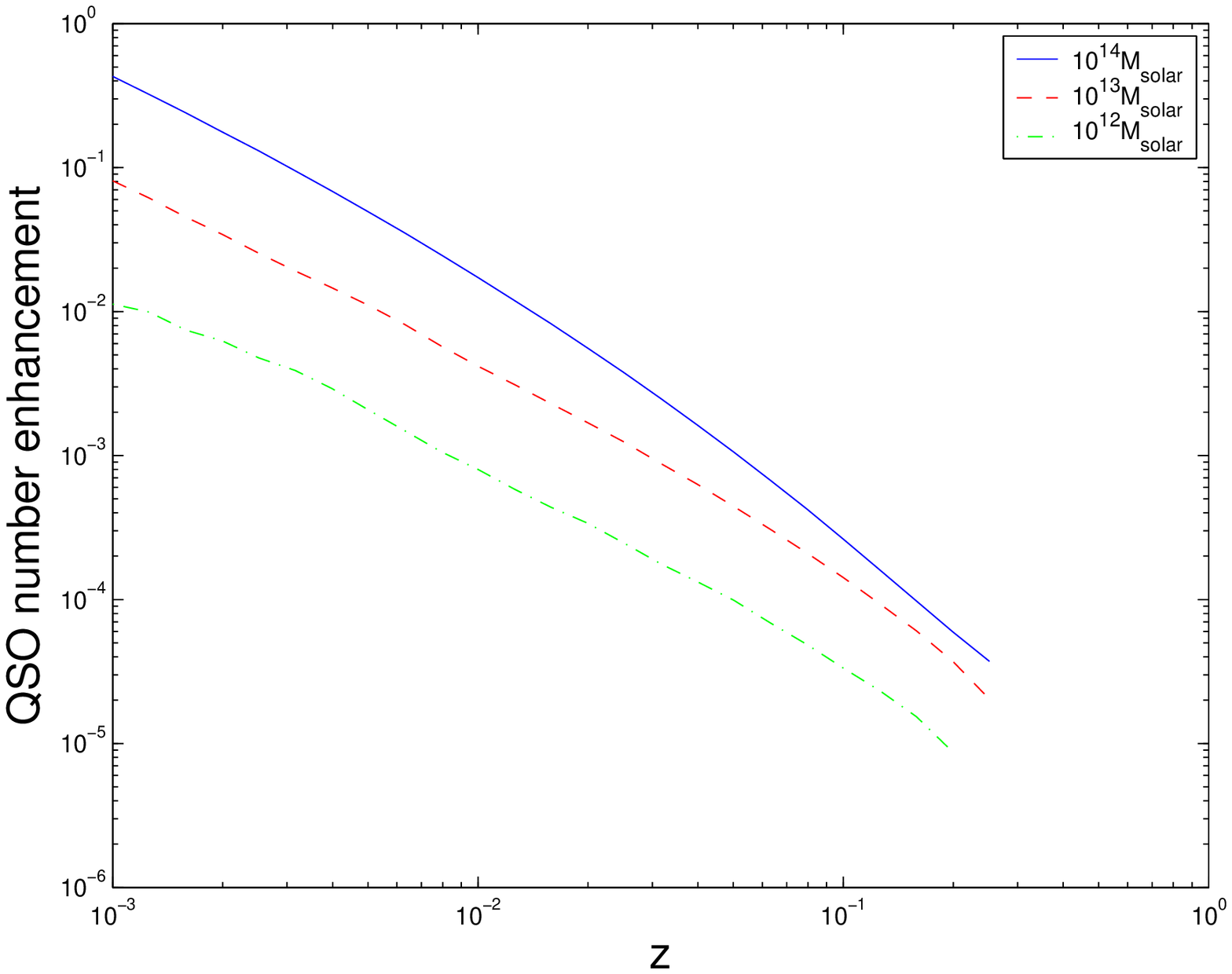}
  \vspace{-5mm}
  \caption{{\small Microlensing induced QSO number enhancement in the field of a galaxy within a radius 20 kpc at different redshifts, and the three curves denote different total masses of microlenses in the galaxy as $10^{14} M_\odot$, $10^{13}
M_\odot$, and $10^{12} M_\odot$ respectively.} }
  \end{minipage}%
  \begin{minipage}[t]{0.5\textwidth}
  \centering
  \includegraphics[width=65mm,height=60mm]{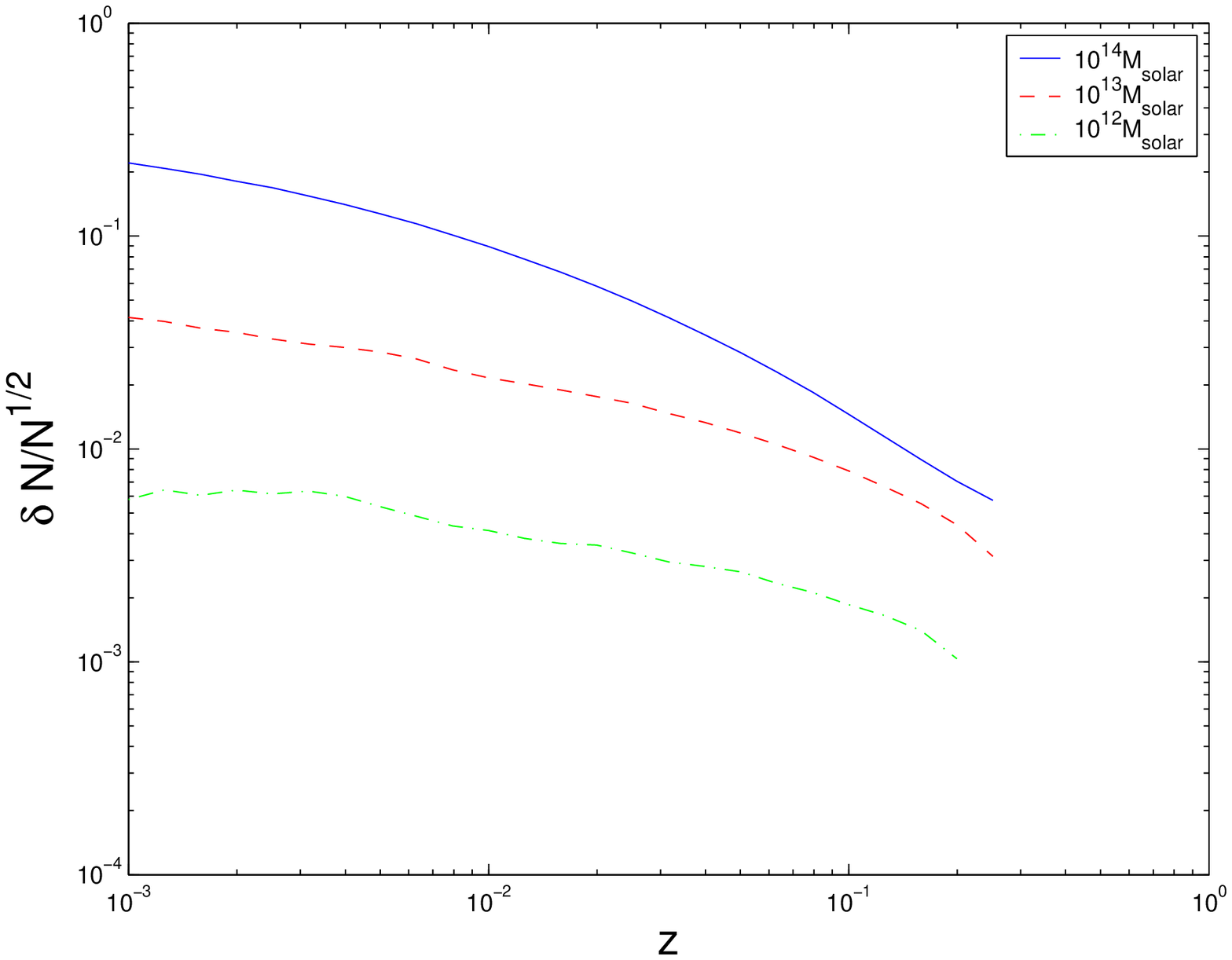}
  \vspace{-5mm}
  \caption{{\small Microlensing induced QSO number enhancement over statistical uncertainty. Parameters are the same as in Fig. 1.}}
  \end{minipage}%
  \label{Fig:fig12}
\end{figure}

\section{Application to NGC 3628}
\label{sect:NGC 3628}
NGC 3628 is a well-studied nearby edge-on Sbc peculiar galaxy in the Leo Triplet. Dahlem et al. (1996) have listed
many X-ray sources in the ROSAT-detected hot gaseous halo of NGC 3628, and they found this number density is
higher than background with a $1.5\sigma$ deviation. They also pointed out that most of these X-ray sources in the
halo of NGC 3628 are probably background AGNs. Recently, Arp et al. (2002) identified several confirmed and
probable QSOs in the halo of NGC 3628 and they found these QSOs are highly concentrated around NGC 3628. Confirmed
X-ray QSOs from Weedman (1985), Dahlem et al. (1996) and Arp et al. (2002) are listed in Table~\ref{Tab:data}.

\begin{table}[h]
  \caption[]{Confirmed X-ray QSOs near NGC 3628. These are taken from the whole-sky X-ray/radio/optical
  verlays catalogue by Flesch, which is accessible at ftp://quasars.org/quasars. }
  \label{Tab:data}
  \begin{center}\begin{tabular}{clccll}
  \hline\noalign{\smallskip}
No &   Survey ID & RA(J2000) & Dec & Redshift & Count rate \\
  \hline\noalign{\smallskip}
1  & Wee 48     & 11 19 46.9 & 13 37 59 & 2.06 &2RXP 6cts/hr        \\
2  & Wee 51     & 11 20 11.9 & 13 31 23 & 2.15 &1RXH 5cts/hr        \\
3  & 1WGAJ1120.2+1332     & 11 20 14.7 & 13 32 28 & 0.995 &1RXH 3cts/hr        \\
4  & 1WGAJ1120.4+1340     & 11 20 26.2 & 13 40 24 & 0.981 &1RXH 9cts/hr        \\
5  & 1WGAJ1120.6+1336     & 11 20 39.9 & 13 36 20 & 0.408 &1RXH 3cts/hr        \\
6  & Wee 52     & 11 20 41.6 & 13 35 51 & 2.43 &1RXH 3cts/hr        \\
7  & Wee 55     & 11 21 06.1 & 13 38 25 & 1.94 &2RXP 22cts/hr        \\
  \noalign{\smallskip}\hline
  \end{tabular}\end{center}

\end{table}

The number density of QSOs in the fields of NGC 3628 is shown in Fig.3, which is about three time the number
density of background. Fig. 4 show the variation of net enhancement factor $q$ of QSO number with a magnification
$M$ for several observational flux limit $f_0$, which is given by:

\begin{equation}
q(M, f_0)=\frac{1}{M} \frac {N(>f_0 /M)}{N(>f_0)}
\end{equation}

For NGC 3628, the flux limit is about $3 \times 10^{14} erg/s$, and the net enhancement factor is about $3$.
Therefore, if present understanding of luminosity function of QSO is right, such concentration cannot be the
result of gravitational lensing. However, as pointed out by Narayan (1989), since the net enhancement of a given
flux limit is dominated by those QSOs with flux larger than $f_0/M$ and less than $f_0$, if the QSO counts are
steeper at faint magnitudes than present understanding, higher concentration could be caused by gravitational
lensing.

\begin{figure}[h]
  \begin{minipage}[t]{0.5\linewidth}
  \centering
  \includegraphics[width=65mm,height=50mm]{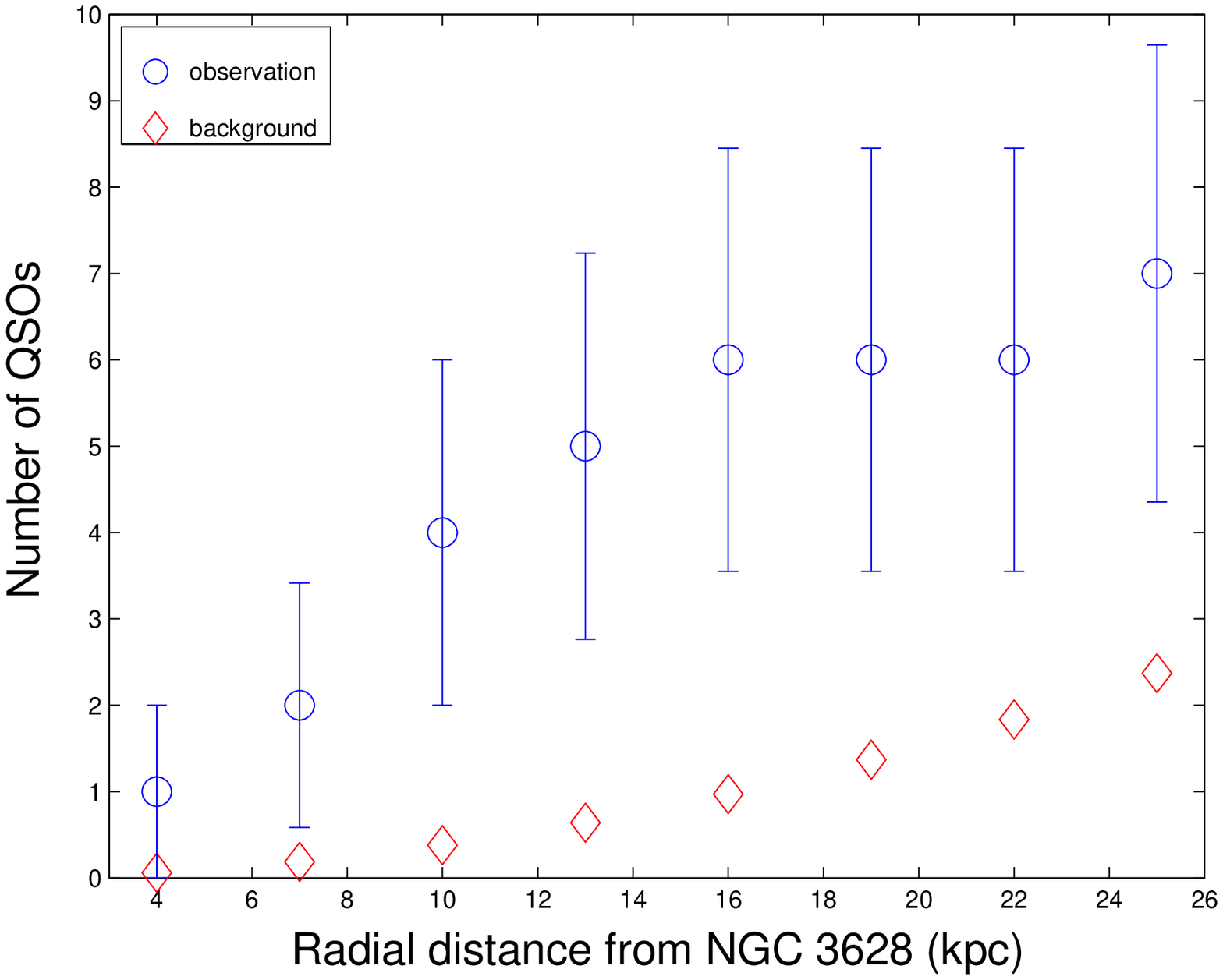}
  \vspace{-5mm}
  \caption{{\small The number of confirmed X-ray selected QSOs within a radial distance from NGC 3628 (circles), compared with the expected number of X-ray selected background QSOs (diamonds).
  } }
  \end{minipage}%
  \begin{minipage}[t]{0.5\textwidth}
  \centering
  \includegraphics[width=65mm,height=50mm]{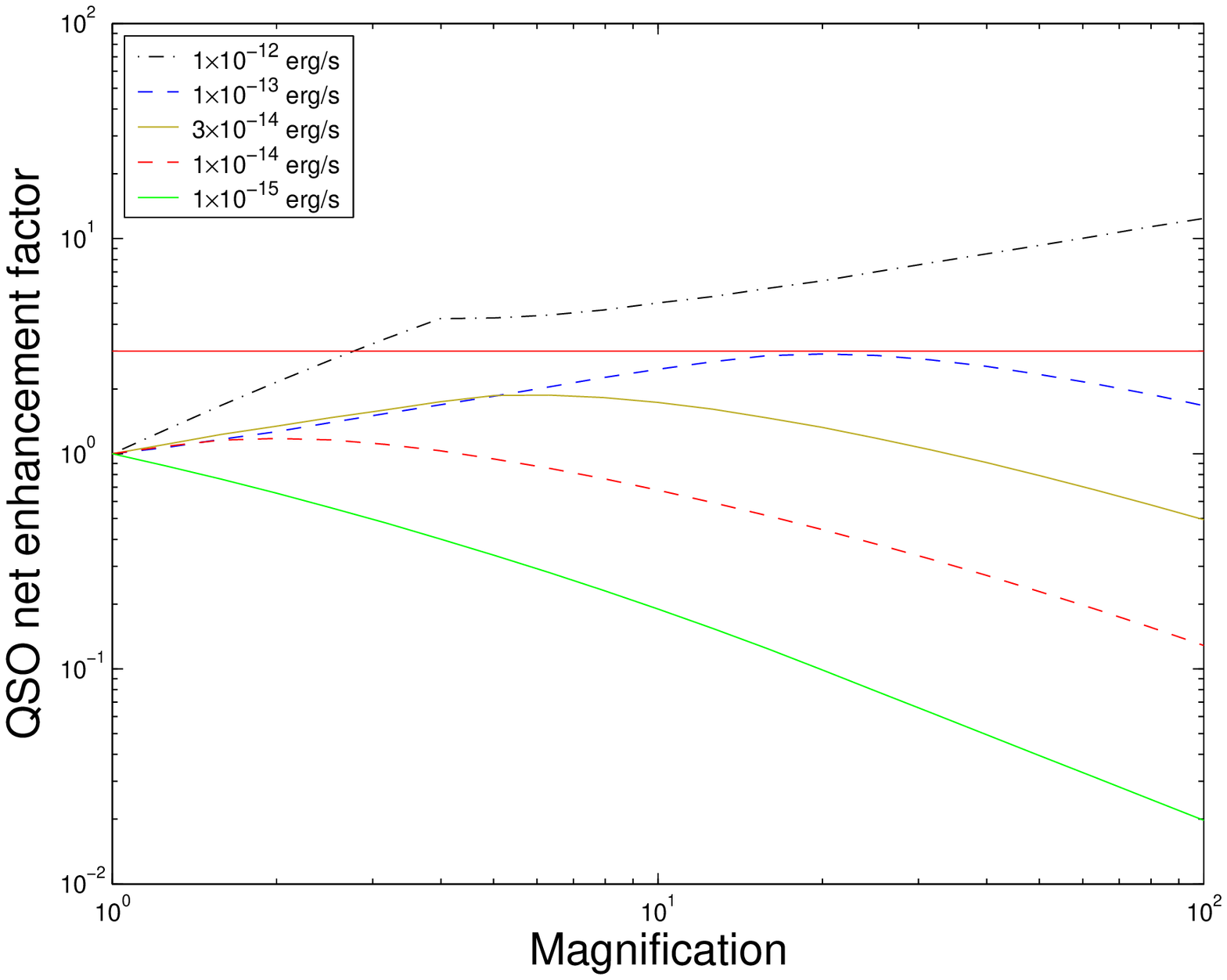}
  \vspace{-5mm}
  \caption{{\small Variation of QSO number density enhancement with magnification, for five limiting fluxes. The red solid line represents the enhancement factor of 3 observed in NGC 3628.}}
  \end{minipage}%
  \label{Fig:fig34}
\end{figure}

\section{Conclusions and discussions}
\label{sect:conclusion}
We have reached the following conclusions in this paper:

1.  From eqn. (4), microlensing (e.g. from black holes)  in nearby
foreground galaxies may enhance the number of background QSOs in
flux limited detection much more effectively than lensing effects
caused by smooth lenses (e.g. WIMPs).

2.  From Fig. 1, the QSO number enhancement is sensitive to the
distance of the foreground galaxy, and thus may be used as a
distance measure of galaxies.

3.  The observed QSO excess in NGC 3628 cannot be caused by gravitational lensing, unless the QSO counts are
steeper at faint magnitudes than present understanding.

However, the following point is worth further discussions. For point mass lenses, since the area within the
Einstein radius is proportional to the total mass, amplification is independent of the mass function of lenses
when the optical depth is low. However, as pointed out by Paczynski (1986a), a point source approximation is not
suitable for microlensing on objects with low mass such as less massive than Jupiter, when the projection of the
QSO onto the sky is larger than the Einstein rings of point mass lenses. Therefore if lenses consist of very low
mass objects, microlensing effect will be less
important. \\

\begin{acknowledgements}
We thank the Dr.X.P. Wu for interesting discussions. The referee Dr. R. Soria and Dr. S. D. Mao are acknowledged
for many comments and suggestions, which allowed us to clarify several points and improve the readability of the
paper. This study is supported in part by the Special Funds for Major State Basic Research Projects and by the
National Natural Science Foundation of China (project no.10233030).
\end{acknowledgements}

\label{lastpage}


\begin{thebibliography}{99}
  \bibitem[2003]{Afonso03} Afonso, C. et al. 2003, \aap, 400, 951

  \bibitem[2001]{Alcock01} Alcock, C. et al. 2001, \apj, 550, L169

  \bibitem[2002]{Arp02} Arp, H. C., Burbidge, E. M., Chu, Y., Flesch, E., Patat, F., \& Rupprecht, G. 2002, \aap, 391,
  833

 \bibitem[2004]{Arp04} Arp, H., Gutierrez, C. M., \& Lopez-Corredoira,
 M., A\&A, submitted (astro-ph/0401103)

  \bibitem[2001]{Benitez01} Benitez, N., Sanz, J. L., \& Gonzalez, E. M. 2001, MNRAS, 320, 241

  \bibitem[2003]{Bennett03} Bennett, C. L. et al. 2003, ApJS, 148, 1

  \bibitem[2003]{Burbidge03} Burbidge, G., Burbidge, E. M., \& Arp, H. C. 2003a, \aap, 400, L17

  \bibitem[2003]{Burbidge03} Burbidge, E. M., Burbidge, G., Arp, H. C., \& Zibetti, S. 2003b, \apj, 591, 690

  \bibitem[1998]{Cole98} Cole, G. H. J., Mundell, C. G., \& Pedlar, A. 1998, MNRAS, 300, 656

  \bibitem[1996]{Dehlem96} Dahlem, M., Heckman, T. M., Fabbiano, G., Lehnert, M. D., \& Gilmore, D. 1996, \apj, 461, 724

  \bibitem[1975]{de Vaucouleurs75} de Vaucouleurs G., 1975, in Sandage A., Sandage M, Kristian J., eds, Stars and
  Stellar Sysems, Vol. 9, Galaxies and the Universe. Univ. Chicago Press, Chicago, p. 309

  \bibitem[2000]{Lasserre00} Lasserre, T., et al. 2000, \aap, 355, L39

  \bibitem[1985]{Lacey85} Lacey, G. C., Ostriker, J. P. 1985, \apj, 299, 633

  \bibitem[1999]{Mao99} Mao, S. 1999, Invited review for ``Gravitational Lensing: Recent Progress and Future Goals", Boston University, July 1999, ed. T.G. Brainerd and C.S. Kochanek
  (astro-ph/9909302)

 \bibitem[2004]{miller2004} Miller, M.C., Colbert, E.J.M., 2004, Int.J.Mod.Phys., D13, 1

  \bibitem[2004]{Muno04} Muno, M. P., Baganoff, F. K. et al. 2004, ApJ, in press (astro-ph/0402087)

  \bibitem[2000]{Murali00} Murali, C., Arras, P., \& Wasserman, I. 2000, MNRAS, 313, 87

  \bibitem[1989]{Narayan89} Narayan, R. 1989, \apj, 339, L53

  \bibitem[2003]{Negi03} Negi, P. S. 2003, Grav. \& Cosmol., Vol. 9, No. 4, 291 (astro-ph/0405572)

  \bibitem[1984]{Nityananda84} Nityananda, R., Ostriker, J. P. 1984, JApA, 5, 235

\bibitem[2004]{overduin04}Overduin, J.M., Wesson, P.S., 2004, Physics Report, in press (astro-ph/0407207)

  \bibitem[1986]{Paczynski86} Paczynski, B. 1986a, \apj, 301, 503

    \bibitem[1986]{Paczynski86} Paczynski, B. 1986b, \apj, 304, 1

   \bibitem[2003]{Romanowsky03} Romanowsky, A.J. et al. 2003, Science, 301, 1696

  \bibitem[1987]{Snchneider87a} Schneider, P. 1987a, \aap, 179, 71

  \bibitem[1987]{Snchneider87b} Schneider, P. 1987b, \aap, 179, 80

  \bibitem[2001]{Sofue01} Sofue, Y., Rubin, V. 2001, ARA\&A, 39, 137

  \bibitem[1984]{Turner84} Turner, E. L., Ostriker, J. P., \& Gott, J. R. 1984, \apj, 284, 1

  \bibitem[1985]{Weedman85} Weedman, D. 1985, ApJS, 57, 523

  \bibitem[1993]{Wilding93} Wilding, T., Alexander, P., \& Green, D. A. 1993, MNRAS, 263, 1075

  \bibitem[2000]{Williams00} Williams, L. R. 2000,\apj, 535,37

  \bibitem[2003]{Ueda03} Ueda, Y., Akiyama, M., Ohta, K., \& Miyaji, T. 2003, \apj, 598, 886

  \bibitem[2004]{Yoo04} Yoo, J., Chaname, J., \& Gould, A. 2004, \apj, 601, 311

  \end{thebibliography}
\end{document}